\documentclass[11pt,a4paper,openany]{article}
\usepackage{amsmath,amssymb,amssymb}
\usepackage{bm}
\usepackage[nottoc]{tocbibind}
\usepackage[dvipdfmx]{color}
\usepackage[dvipdfmx]{graphicx}
\usepackage{ascmac}
\usepackage{indentfirst}
\usepackage{docmute}
\usepackage{booktabs}
\usepackage{fancyhdr}
\usepackage{geometry} 
\usepackage{braket}
\usepackage{subfig}
\usepackage{caption}
\usepackage {colortbl,array,xcolor}

\usepackage{here}
\usepackage{cite}
\usepackage{setspace}
\begin{document}
%title
\title{A single electron-photon pair generation\\ from a single polarization-entangled photon pair}
\author{Kazuyuki Kuroyama$^{1,a)}$, Marcus Larsson$^{1)}$, Sadashige Matsuo$^{1)}$, Takafumi Fujita$^{1)}$, \\Sascha R. Valentin$^{2)}$, Arne Ludwig$^{2)}$, Andreas D. Wieck$^{2)}$, Akira Oiwa$^{3)}$ and Seigo Tarucha$^{1,4)}$}
\setcounter{page}{1}
\date{}
\maketitle
\thanks{a) Electronic mail: kuroyama@meso.t.u-tokyo.ac.jp}
\newline
\newline

\thanks{\noindent 1 Department of Applied Physics, The University of Tokyo, Bunkyo-ku, Tokyo, Japan\\
 2 Lehrstuhl f\"{u}r Angewandte Festk\"{o}rperphysik, Ruhr-Universit\"{a}t, Bochum, Germany\\
 3 The Institute of Scientific and Industrial Research, Osaka University, Ibaraki, Osaka, Japan\\
 4 Center for Emergent Materials Science, RIKEN, Wako, Saitama, Japan } 

\section*{Abstract}
We demonstrate paired generation of a single photo-electron in a quantum dot and a single photon from a single polarization-entangled photon pair that is generated by spontaneous parametric down conversion.  The electron is created in a GaAs quantum dot via inter-band resonant excitation by irradiating one of the paired photons on the dot, while the remaining photon is detected by a single photon counter. Because the paired photons are generated simultaneously, the photo-electron and the remaining photon should be observed as a coincident event if the traveling distance is equivalent between them. The coincidence is probabilistic due to the probabilistic generation of the paired photon. We observe the probability of finding the photon with the photo-electron detection substantially higher then that without the photo-electron detection and derive the coincident rate comparable to the emission rate of the original photon pairs. 
%%%%%%%%%%%
\section*{Background and overview of this work}
Quantum entanglement between different forms of qubits is an indication of universality of quantum mechanics and especially entanglement between a solid-state qubit and a photonic qubit is an important ingredient to realize distributed photonic quantum communication based on quantum repeaters \cite{gobby2004quantum,PhysRevLett.81.5932}. Experimental demonstrations of the solid-state qubit-photonic qubit entanglement have been reported in a few systems: A hole spin in an InGaAs self-assembled quantum dot \cite{gao2012observation,de2012quantum}, an electron spin confined in a IaAs quantum dot \cite{PhysRevLett.110.167401} and a nitrogen vacancy center (NV center) in diamond \cite{togan2010quantum}. These qubits may have difficulty in the scale-up and complex quantum gating both needed to combine the quantum communication with gate-based computation. On the other hand electron spins in GaAs lateral quantum dots (QDs) is another candidate for generating the solid-state qubit-photon qubit entanglement. The GaAs QD has advantages compared to the other solid-state qubits in the ability of quantum gating\cite{PhysRevLett.113.267601,malinowski2016notch} and scale-up\cite{PhysRevB.90.075436,doi:10.1063/1.4869108,doi:10.1063/1.4875909,ito2016detection}. Quantum state transfer from a photon polarization to an electron spin \cite{Vrijen2001PhysicaE:Low-dimensionalSystemsandNanostructures} was experimentally demonstrated for ensembles of photons and those of electrons with a GaAs quantum well \cite{PhysRevLett.100.096602}. For single photons and single electrons we previously demonstrated angular momentum transfer from the single photon polarization to the single electron spin using GaAs QDs \cite{Fujita2015}. Furthermore, single polarization-entangled photon pairs can be used to establish the photon-spin entanglement if the spin is generated in the QD by one of the paired photons. However, no experiment has yet been demonstrated on photo-electron generation in QDs using single entangled photon pairs. 
\\	\ \
In this paper, we report on the experimental observation of the coincident generation of a photo-electron in a GaAs QD and a photon from single polarization-entangled photon pairs. One of the paired photons is irradiated on the GaAs QD, while the other on a single photon counter. Real-time measurements of the photo-electron in the dot by a charge sensor \cite{PhysRevLett.106.146804,PhysRevLett.110.266803,PhysRevB.90.085306} and the photon by the photon counter are performed simultaneously to derive the temporal coincidence between them. The averaged interval of single photons irradiated on the dot and the photon counter is adjusted to be larger than the time resolution of the charge sensor but much larger than that of the photon counter. Therefore zero to two photons (N=0 to 2) are mostly detected by the photon counter within the response time of detecting a photo-electron by the charge sensor. However, this does not always indicate the photo-electron-photon temporal coincidence arising from the entangled photon pairs, because the generation efficiency of the entangled photon pairs is only 10 to 15 \%. We compare the probability of one or two photon detection with and without the photo-electron detection and finally confirm the coincident generation of a photo-electron in the dot and a photon detected by the photon counter.
%%%%%%%%%%
\section*{Optical setup for SPDC} 
Figure 1 schematically shows the experimental setup consisting of two sections: Section I for generating single photon pairs using a spontaneous parametric down conversion (SPDC) technique \cite{PhysRevLett.75.4337,PhysRevA.60.R773}and Section II for detecting photo-electrons in the dot and photons by the photon counter upon separate irradiation of single photons from the photon pairs. 
\\ \ \
A Ti:sapphire ultra-short pulsed laser (MIRA-900P) whose auto correlation width is 3 psec and repetition rate is 76 MHz is used to generate the SPDC photons. The laser pulse has a peak power of 2 kW with the center wavelength of 808 nm. This wavelength is adjusted to coincide with excitation energy for the heavy hole exciton in the GaAs QD which is determined from a spectrum of photo-electron trapping efficiency in the dot (see SI). The photon energy is doubled using second harmonic generation (SHG) with a Type-I Beta-Barium-Borate (BBO) crystal. To extract only the SHG photons a Pellin-Broca prism, several irises and short pass filters are used. After this energy cleaning, the average power of the second harmonic light obtained is about 100 mW. Then, the SHG photons are irradiated on a Type-II BBO crystal to generate SPDC photon pairs. A band pass filter whose transmittance is maximal at 800 nm is placed just after the crystal to only transmit the SPDC photon pairs. The paired photons have two kinds of linear polarization orthogonal to each other and are emitted in two different directions with mirror symmetry. The inset of Fig.2 shows the far-field image of the emitted photons observed by a CCD camera. The upper, and lower ring indicates the horizontally, and vertically polarized photons, respectively. The polarization of the photons coming on the crossing points of the two rings cannot be distinguished, and therefore the paired photon state is polarization-entangled as represented by
$(\ket{HV}+\ket{VH})/\sqrt{2}$, where $\ket{H}$ and $\ket{V}$ indicates the horizontally, and vertically linearly polarized photon state, respectively. A series of two irises are placed on each crossing point behind the band pass filter to extract the paired photons entangled but emitted in two directions, and the photons are separately coupled into two multimode fibers whose core diameter is 50 $\rm{}\mu$m. Here we call the photon path through the right (left) crossing point of the two photon rings as Path A (B).  
\\ \ \	
	In order to confirm the polarization correlation of the paired photons we executed coincidence measurement between the optical fiber outputs. Each output is detected by an avalanche photodiode mounted on a single photon counting module (SPCM). The SPCM signal synchronized with repeating pulses of the Ti: sapphire laser is counted using a PCI board (Time Harp 200, Pico Quanta). To selectively pick up the coincidence signals, a logical AND is performed on the two different SPCM signals or two optical fiber outputs. From the polarization correlation measurement taken on the coincidence photons (see SI), we confirmed that the detection rate of the entangled photon pairs is 8 to 10 kHz (10 to 15 \% of all detected photons by SPCM).
%%%%%%%%%%
\section*{GaAs lateral quantum dot}
The GaAs QD device used for trapping single photons is defined by the surface Schottky gates  in a two-dimensional electron gas formed in a 15 nm thick GaAs well located between two thick $\rm{}Al_{0.33}Ga_{0.67}As$ barriers (see Fig. 3(a) and (b)). To enhance the efficiency of photon absorption in the dot, a distributed Bragg reflector (DBR) consisting of an $\rm{}Al_{0.10}Ga_{0.90}As/AlAs$ multiple layer is embedded. For the photon trapping experiment a single QD is formed by applying appropriate voltages to the gates. On both side of the dot another dot is placed as a charge sensor. The right charge sensor is only used in the experiment. In addition a 300 nm thick Au/Ti mask (not seen in Fig. 3(b)) having a 500 nm-diameter aperture is placed on top of the photon-trapping QD to avoid photon irradiation on the device except for the dot. All of the measurement is carried out for the dot device placed in a dilution fridge (TRITON 200) with base temperature of 25 mK. The photons are irradiated on the dot through an optical window equipped at the bottom of the fridge. Note an electron-heavy hole pair is photo-excited but the hole is quickly ejected from the dot because of the applied negative gate voltages while leaving the electron in the dot. 
\\ \ \
The single photo-generated electron is detected by measuring the QD charge sensor response upon the photon irradiation. A change in the number of electrons in the dot modifies the charge sensor current, $\rm{}I_{sensor}$, which is measured using a radio frequency reflectometry technique called rf-QD \cite{doi:10.1063/1.2794995}. The minimal charge detection time achieved is 10 $\rm{}\mu$sec given by the time of integrating $\rm{}I_{sensor}$ to achieve an appropriate signal-to-noise ratio. 	
\\ \ \
We adjusted the gate voltages of the single QD such that the electron escape time or photo-electron trapping time is sufficiently longer than the minimum rf-QD detection time. The QD charge state as a function of gate voltages can be determined using the charge sensor as shown in Fig. 2(c), differential conductance d$\rm{}I_{sensor}$/d$\rm{}V_R$ of the charge sensor v.s. two gate voltages $\rm{}V_R$ and $\rm{}V_L$ without the photon irradiation. Sharp lines indicated by the arrows are the charge state transition lines separated by the Coulomb gap and the QD level spacing. The lowest line indicates the transition between the zero and one electron state in the dot, and the number of electrons in the dot increases one-by-one every time when crossing the transition line upward. Figure 2(d) shows the real time trace of $\rm{}I_{sensor}$ measured at the bias point on the transition line. Transitions between the two current values indicate the charge tunneling between the dot and the source or drain electrode. The time scale of electron trapping can be evaluated from the time interval needed for $\rm{}I_{sensor}$ switching from low to high.
\\ \ \ 
In the photon irradiation experiment we adjusted the electron trapping time to a few hundred micro seconds about ten times longer than the averaged interval of the down converted photons. The charge state is tuned to be in the Coulomb blockade region about 5 mV below the third charge transition line in the direction of $\rm{}V_L$  marked by the green circle in Fig. 2(c). 
%%%%%%%%%%
\section*{Coincidence measurement on the electron and the photon}
Next, we generated photon pairs and then irradiated one of the paired photons on the dot. Before the photo-electron photon coincidence measurement we readjusted the single photon count rates for the photons along the two paths, Path A and B, using two SPCMs and found the coincidence count rate between them was 8 to 10 kHz, which corresponds to 12 to 15 \% of the photon detection rates. The charge sensor signal, $\rm{}I_{sensor}$ and the photon counter signal, $\rm{}V_{photon}$ are both monitored using a digitizer whose data acquisition rate is 100 MHz. The integration time of $\rm{}I_{sensor}$, and $\rm{}V_{photon}$ is 10 $\rm{}\mu$sec, and 10 nsec, respectively. 
\\ \ \
Figure 3 (a) and(b) show the time trace of $\rm{}I_{sensor}$ together with that of $\rm{}V_{photon}$ measured simultaneously: red circles for $\rm{}I_{sensor}$ and green spikes for $\rm{}V_{photon}$. Each $\rm{}V_{photon}$ spike indicates one photon detection by the single photon counter. The $\rm{}I_{sensor}$ time trace measured upon the photo-electron trapping is similar to that of Fig. 2(d) (see SI). $\rm{}I_{sensor}$ abruptly decreases to the low level when an electron is trapped in the dot and then stays at the low level, indicating that a single photo-electron is trapped by the dot in the 10 $\rm{}\mu$sec time window of orange color. In this time window one, or two $\rm{}V_{photon}$ spikes or N=1 or 2 photons are mostly detected by SPCM in (a), and (b), respectively. The efficiency of the photo-electron trapping by the dot is about 0.1 \% [SI], so we repeated the experiments of separately irradiating paired photons on the dot and on the photon counter, and finally acquired 27 time traces indicating the photo-electron trapping by the dot and calculated the probabiliti $\rm{}P_{coincidence}$(N) of finding N (= 0, 1, 2, … ) photons in the photo-electron trapping time window. The obtained probabilities are $\rm{}P_{coincidence}$(0)=44$\rm{\pm}$9.7 \%, $\rm{}P_{coincidence}$(1)=41$\rm{\pm}$9.5 \%, and $\rm{}P_{coincidence}$(2)=15$\rm{\pm}$6.8 \%. $\rm{}P_{coincidence}$(2) is due to the single photon detection twice in the photo-electron trapping time window, and appears unintentionally because the time window is not small enough compared to the average interval of the photon detection signals. Note the photon detection observed in the photon trapping time window does not always indicate the coincident detection of a photo-electron in the dot and a photon in the photon counter generated from a single polarization-entangled photon pair. To evaluate the probability of the non-coincident detection in $\rm{}P_{coincidence}$(1) and $\rm{}P_{coincidence}$(2) we derived the averaged probability $\rm{}P_{random}$(N) of detecting N=1 or 2 photons per 10 $\rm{}\mu$sec by the photon counter regardless of whether the photo-electron is detected or not. The obtained $\rm{}P_{random}$(1) = 27$\rm{\pm}$1.5 \% and $\rm{}P_{random}$(2) = 8.9$\rm{\pm}$0.95 \% for the N=0, 1 and 2 photons are listed in the lower column of Table1. $\rm{}P_{coincidence}$(1) and $\rm{}P_{coincidence}$(2) are both significantly greater than $\rm{}P_{random}$(1) and $\rm{}P_{random}$(2) due to the true coincident detection of a photon and a photo-electron generated from the single entangled photon pair. From this result we evaluate that about 12 \% of the photo-electron trapping events should have the true coincidence photons detected by SPCM [SI]. This percentage indeed compares well to that of the polarization-entangled photon pairs detected by two SPCMs as described before. 
%%%%%%%%%%
\section*{Further consideration about the result}
Finally we consider a way to improve the coincidence probability. Recent technical advances in rf-QDs charge sensing enable to shorten the charge detection time down to 100 nsec \cite{barthel2010fast}. If the charge detection time decreased to 1 $\rm{}\mu$sec, $\rm{}P_{random}(1)$ can be 1/10 of the current value, while the probability of the true coincident events is unchanged. Then $\rm{}P_{coincidence}(1)$ for the true coincidence becomes 10 times higher than $\rm{}P_{random}(1)$. To increase the generation rate of the polarization-entangled photon pairs will also be efficient to raise the probability of finding the true coincident pairs of the photo-electron and the photon.
%%%%%%%%%%
\section*{Summary}
In summary we performed the coincidence measurement of single photo-electrons in a quantum dot and single photons using a down converted photon pair. We compared the probability of finding the photon with the photo-electron detection with and without the photon detection, and found the former substantially higher than the latter. From this comparison we estimated the coincidence rate to be comparable to the emission rate of the original photon pairs. We previously demonstrated that single electron spins are photo-generated in QDs by irradiating single photons with preserving the angular momentum. Consequently combining with the present result, we here assume that the angular momentum correlation can be preserved between the single photo-electrons in the dot and the single photons that are spatially separated but generated from the single polarization-entangled photon pairs. 
%%%%%%%%%%
\section*{Acknowledgement}
We greatly thank S. Takeuchi and R. Okamoto for important technical advice about the SPDC setup.
This work was partially supported by Grant-in-Aid for Young Scientific Research (A) (No. JP15H05407), Grant-in-Aid for Scientific Research (A) (No. JP16H02204, No. JP25246005), Grant-in-Aid for Scientific Research (S) (No. JP26220710), JSPS Research Fellowship for Young Scientists (No. JP14J10600), JSPS Program for Leading Graduate Schools (ALPS) from JSPS, Japan Society for the Promotion of Science (JSPS)　Postdoctoral Fellowship for Research Abroad Grant-in-Aid for Scientific Research on Innovative Area, "Nano Spin Conversion Science" (No.JP15H01012, No.26103004), Grant-in-Aid for Scientific Research on Innovative Area, "Topological Materials Science" (Grant No. JP16H00984) from MEXT, CREST, and the Murata Science Foundation.
%%%%%%%%%%\section*{Figures}

%figure 1
%%
\begin{figure}[H]
\begin{center}
\includegraphics[width=\textwidth]{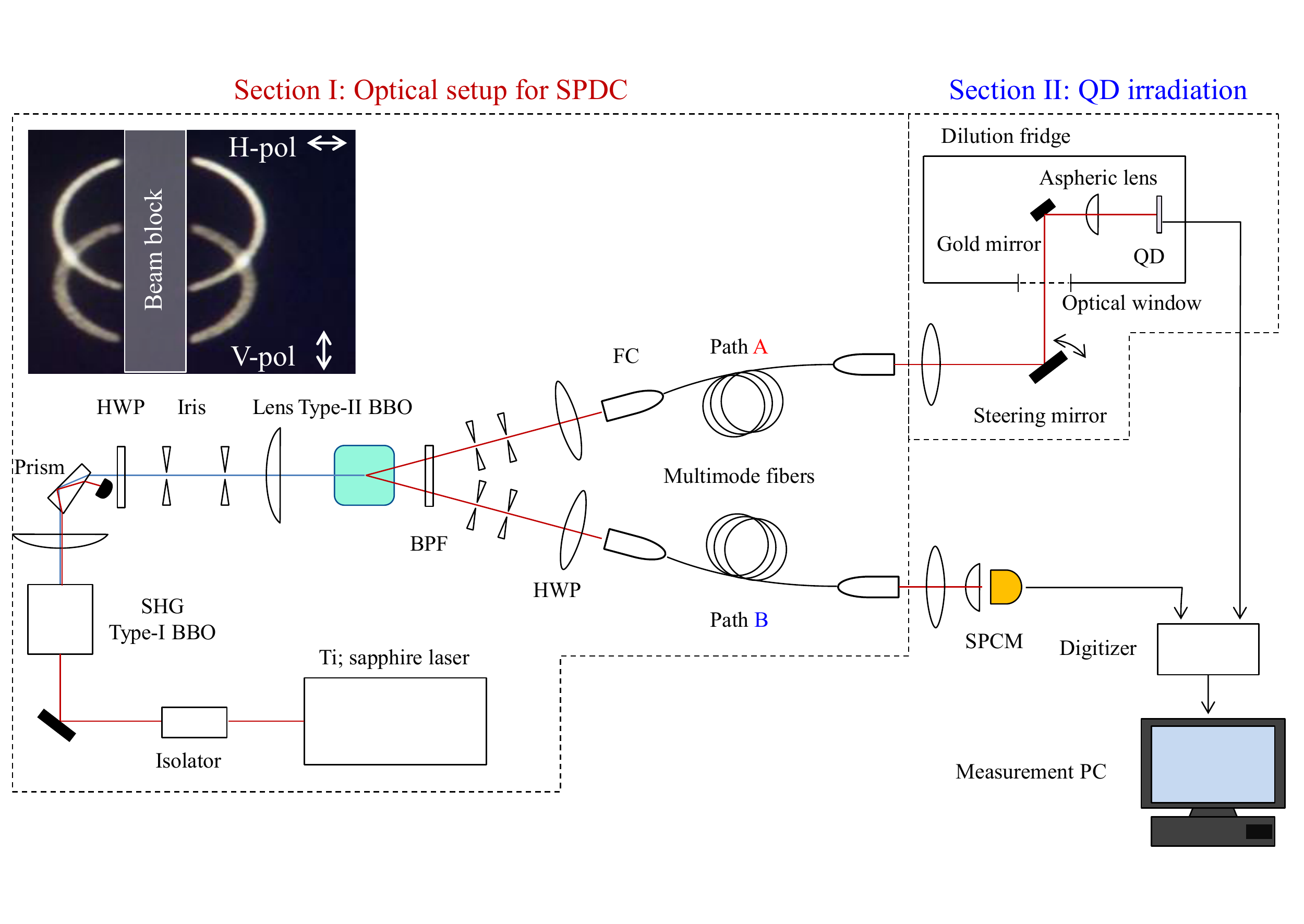}
\caption*{Fig. 1 Schematic of the experiment setup used for the coincidence measurement of the photo-electron excited in the dot and the photon at SPCM. The setup can be separated into two sections, I and II. Section I is an optical setup to generate SPDC. A Ti: sapphire ultra-short pulsed laser is used to generate SHG photons with a Type-I BBO crystal. The SHG photons are irradiated on a Type-II BBO crystal to generate SPDC photons after removing the original laser light with a prism and two irises, and rotating the polarization with a half eave plate (HWP) to satisfy the phase matching condition of SPDC. The paired photons are emitted in two different SPDC photon cones through a band pass filter (BPF). A series of irises are placed to cut out the crossing points of the photon cones. After filtering the energy and the spatial position, the photons are coupled into the multimode fibers. The fiber output photons are observed by SPCM. Section II is a setup for irradiating the photons on the dot and detecting the photo-electron. A dilution refrigerator has an optical window at the bottom and the irradiating photons along Path A are focused by an aspheric lens on the sample through the optical window. The position of the irradiating photon spot can be aligned by tilting a steering mirror. Finally, the charge sensing signal of the photo-electron trapping and the SPCM signal of detecting the photons along Path B photon detected by the SPCM are monitored on the digitizer. Inset is a far-field image of the SPDC photons observed by a near infrared sensitive CCD camera. The upper, and the lower ring indicate emission of horizontally polarizing photons, and vertically polarizing photons, respectively. }
%\label{figure1}
\end{center}
\end{figure}
\newpage
%figure 2
%%
\begin{figure}[H]
\begin{center}
\includegraphics[width=\textwidth]{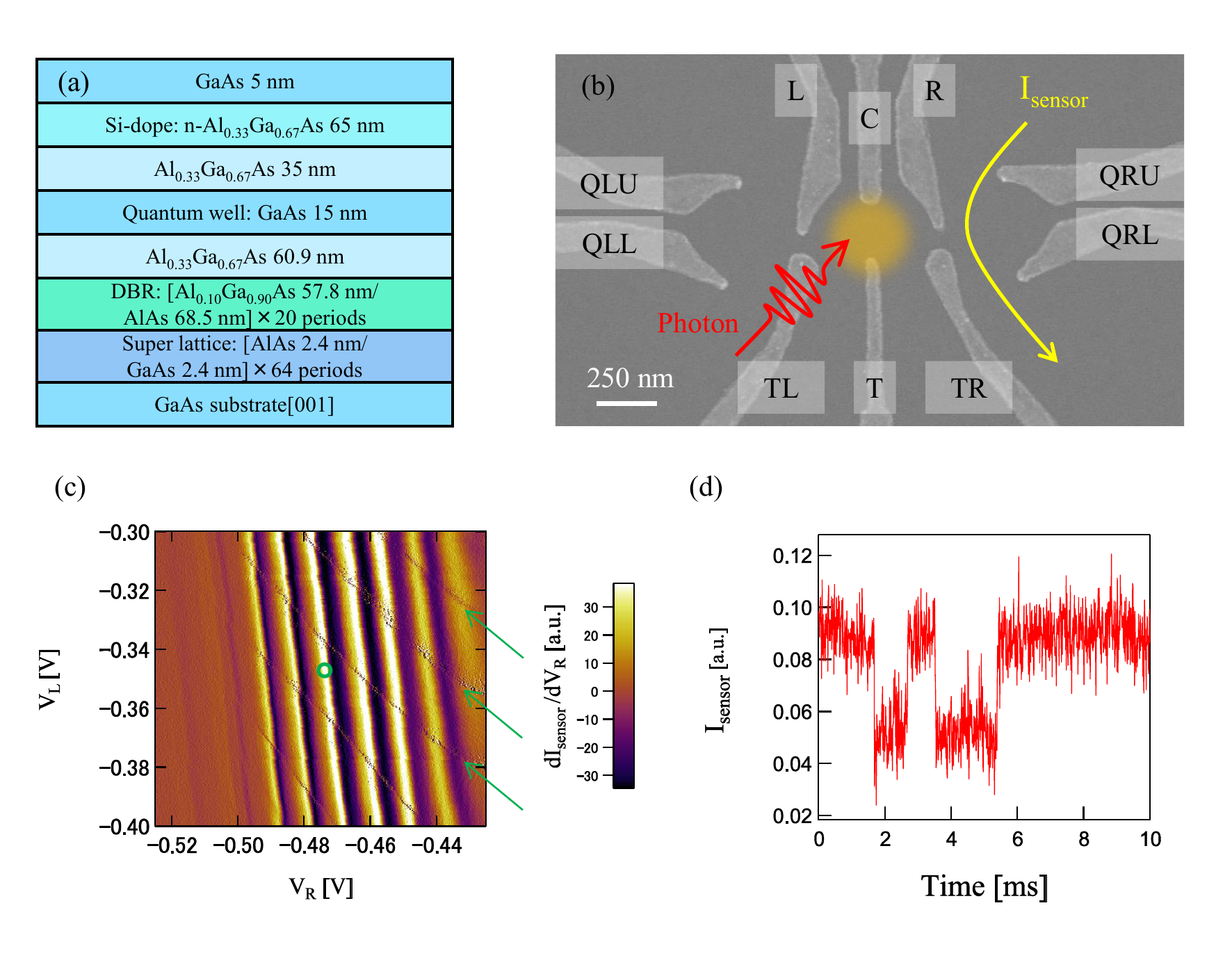}
\caption*{Fig.2 (a) Layer profile of the quantum well wafer used for fabricating the QD device. A 15 nm GaAs well is sandwiched two with barrier layers of ${\rm Al_{0.33}Ga_{0.67}As}$. Below the barrier layer a DBR structure consisting of periodic $\rm{}Al_{0.10}Ga_{0.90}As /AlAs$ quarter-wave stacks is embedded. (b) Scanning electron micrograph of a test device fabricated in the same way. The dot position formed by gating is shown by the orange circle. On the right side of the dot, another dot is formed as a charge sensor and its conduction is measured by a rf circuit. (c) Typical stability diagram of the single dot characterized by the differential conductance of the charge sensor current, d$\rm{}I_{sensor}/dV_{R}$ as a function of two gate voltages $\rm{}V_{L}$ and $\rm{}V_{R}$. The thin gray lines indicated by the green arrows are the charge transition lines. We set the gate voltages $\rm{}V_{L}$ and $\rm{}V_{R}$ at the green circle position between the charge transition lines for the photon irradiation experiment. (d) Typical time trace of the single electron tunneling between the dot and source-drain electrodes. When the electron enters the dot, Isensor abruptly drops. The time trace is taken over 10 $\rm{}\mu$sec for integrating the Isensor data.}
%\label{figure2}
\end{center}
\end{figure}
\newpage
%figure 3
%%
\begin{figure}[H]
\begin{center}
\includegraphics[width=\textwidth]{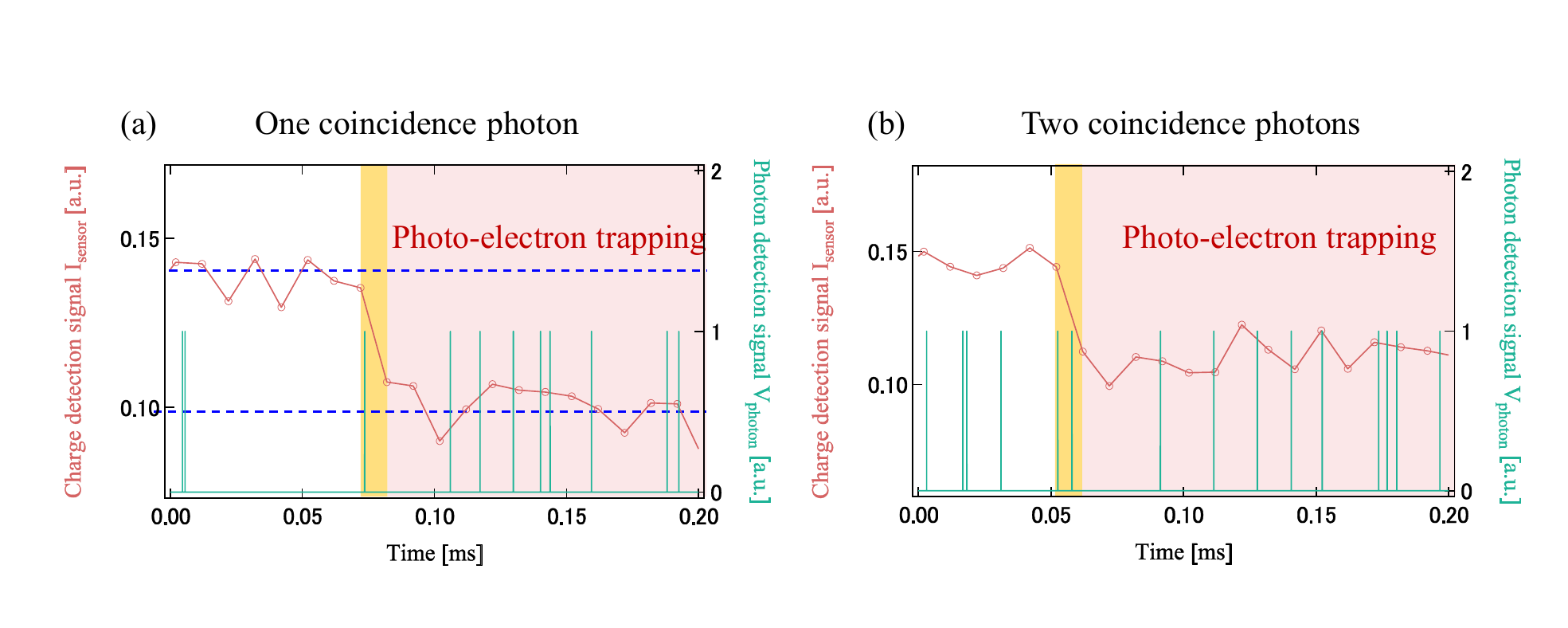}
\caption*{Fig.3 Examples of simultaneously measured time traces for the charge sensor current Isensor in red and the SPCM photon detection signal $\rm{}V_{photon}$ in green. One photon, two photons, and no photon are detected by SPCM in the 10 $\rm{}\mu$sec photon-electron trapping time window in light blue in (a), and (b), respectively. Isensor falls down in the time window, indicating one photon-electron is just trapped by the dot. }
%\label{figure3}
\end{center}
\end{figure}
\newpage
%figure 4
%%

\begin{table}[H]
  \begin{center}
    \begin{tabular}{|c|c|c|c|} \hline
    \rowcolor[gray]{0.8}%
      \shortstack{\\The number of photons \\ detected in the time window} & N=0 & N=1 & N=2 \\ \hline \hline
      \shortstack{\\one electron creation : \\ $\rm{}P_{coincidence}$(N)} & 0.44$\rm{\pm}$0.097 & 0.41$\rm{\pm}$0.095 & 0.15$\rm{\pm}$0.068
 \\ \hline
      \shortstack{\\no electron creation : \\ $\rm{}P_{ramdom}$(N)} & 0.605$\rm{\pm}$0.0162 & 0.273$\rm{\pm}$0.0148 & 0.0893$\rm{\pm}$0.00947
 \\ \hline
    \end{tabular}
    \caption*{Table 1. The Upper column shows the probability of finding N (=0, 1, 2) photons in the photon-electron trapping time window of 10 $\rm{}\mu$sec. The lower is an averaged probability of detecting N=0, 1, 2 photons per 10 $\rm{}\mu$sec by the photon counter regardless of whether the photo-electron is detected or not. The single photon detection twice in the 10 $\rm{}\mu$sec occurs unintentionally because the 10 $\rm{}\mu$sec is not short enough compared to the average interval of the photon detection signals. $\rm{}P_{coincidence}$(1) and $\rm{}P_{coincidence}$(2) are both significantly greater than $\rm{}P_{ramdom}$(1) and $\rm{}P_{ramdom}$(2) due to the true coincident detection of the photon and the photo-electron. }
%\label{figure4}
    \label{tab:price}
  \end{center}
\end{table}

\newpage
%%%%%%%%%%
\section*{Visibility of polarization correlation}
In order to confirm polarization correlation and its visibility on the coincidence photons, we carried out measurement of polarization state of $\ket{H_AV_B}$ and $\ket{V_AH_B}$. The measurement setup is shown in Fig. S1 (a). We put polarization beam splitters (PBSs) on both the path A and the path B to take projection measurement onto the orthogonal linear polarizations. A half wave plate (HWP) is inserted on the path A to observe the visibility of the polarization correlation. we extract only the coincident signals by passing the photon detection signals obtained on the path A and the path B through an AND logical circuit. Results of the projection measurement onto $\ket{H_AH_B}$ (red curve) and $\ket{H_AV_B}$ (blue curve) state are shown in Fig. S1 (b). In case of the $\ket{H_AH_B}$ projection measurement, the coincidence count is maximal when rotating angle of polarization on the path A and the result indicates that the paired photons have the $\ket{V_AH_B}$ state. On the other hand in case of the $\ket{H_AV_B}$ projection measurement, the coincidence count is maximal when rotating angle of the polarization on the path A is 180 degrees and this indicates that the paired photons have a polarization corresponding to the projection basis, $\ket{H_AV_B}$. Based on this we evaluated the visibility of the polarization correlation for the coincident photon pairs, 96 \%. 

\section*{Random sampling}
	As shown in the lower column of the table 1, we took the N photons detection probabilities in the 10 $\rm{}\mu$sec time window. In the result of the random sampling 0 to 4 photons are found in the 10 $\rm{}\mu$sec and an example of the sampling is shown in Fig. S2. The time trace shows cases of the detecting N (=0, 1, 2) photons.
	Photon detection rate on the path B is measured by the SPCM and we saw temporal fluctuation of the photon detection rate. Basically in a case of pulse excitation, the number of photons in 10 $\rm{}\mu$sec should follow Poisson statistics. However, the probability distribution of the random sampling seems not to follow the statistics. One of reasons explaining the disagreement is that the SPCM tends to emit dark signals after detecting a photon.

\section*{Quantum efficiency of the QD device}
	The quantum efficiency of the QD is measured beforehand and is at maximum having about 0.15 \% at the heavy hole resonant excitation. The excitation energy is experimentally determined by the incident photon energy dependence of photo-electron trapping efficiency in the QD shown in the Fig. S3. A higher peak located at 1.534 eV is the heavy hole exciton. The efficiency is evaluated for the incident photons after passing through the optical mask fabricated on the QD.

%figure S1
%%
\begin{figure}[H]
\begin{center}
\includegraphics[width=\textwidth]{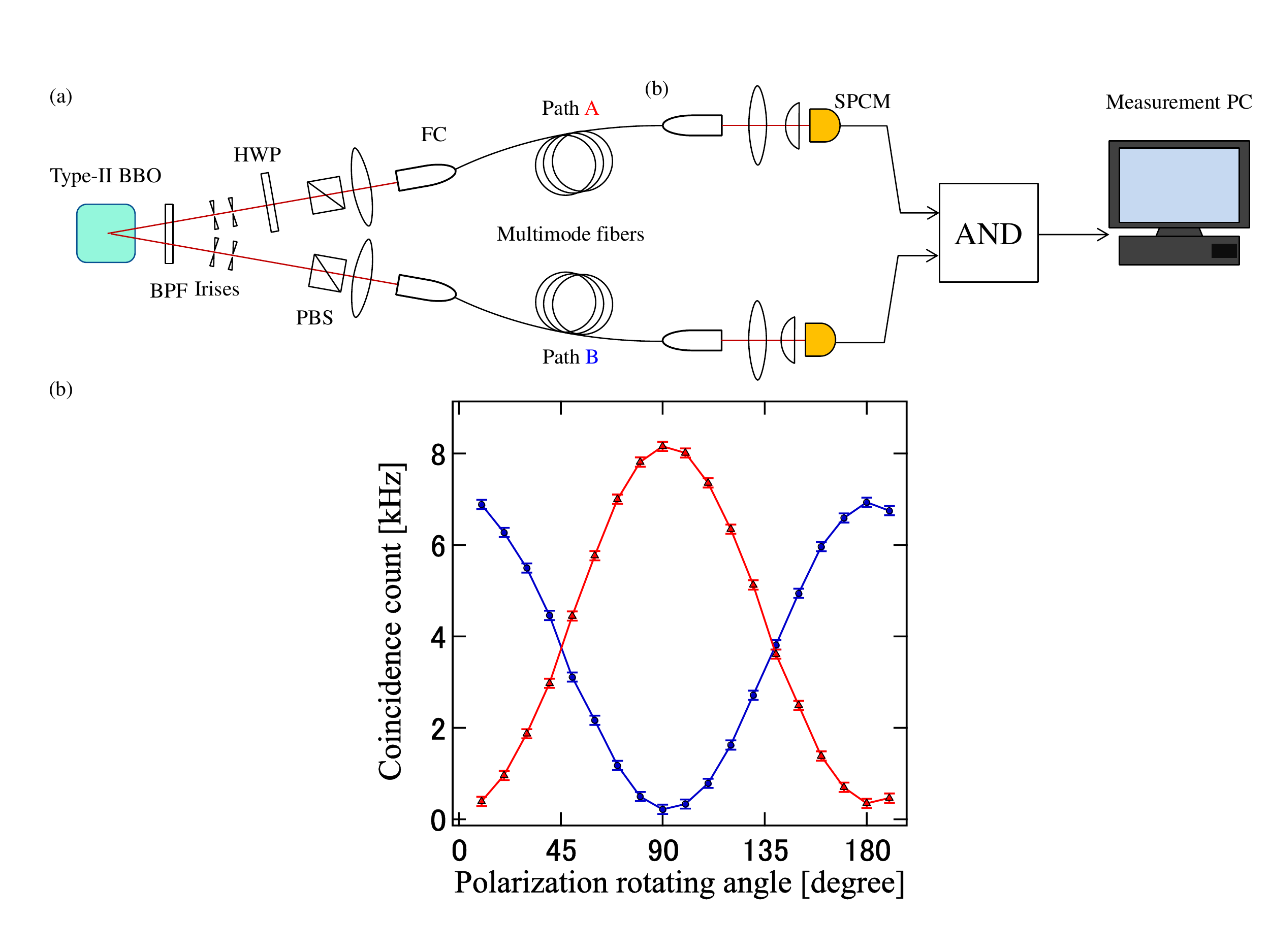}
\caption*{Fig.S1(a) A schematic of an optical setup for the polarization correlation measurement taken on the down converted photons. Polarization state of the photon on the path A is rotated by passing through the HWP. PBSs are used to take the projection measurement onto $\ket{H_AH_B}$ or $\ket{H_AV_B}$ state.  We take a logical AND on the photon detection signals coming from the different SPCMs to extract the coincidence photons. (b) Polarization correlation measured on the simultaneously detected paired photons. A red curve and a blue one are taken by the projection measurement on the $\ket{H_AH_B}$ state and on the $\ket{H_AV_B}$ state, respectively. For the $\ket{H_AH_B}$ projection when the polarization of the path A is rotated 90 degree,  (i.e. the projection is done onto the $\ket{V_AH_B}$ state) the coincidence count is maxamal. On the other hand, for the $\ket{H_AV_B}$ projection, when the polarization is rotated 180 degree(i.e. the projection is done onto the $\ket{H_AV_B}$ , the original state) maximal coincidence is obtained.}
%\label{figureS1}
\end{center}
\end{figure}
\newpage

%figure S2
%%
\begin{figure}[H]
\begin{center}
\includegraphics[width=10cm]{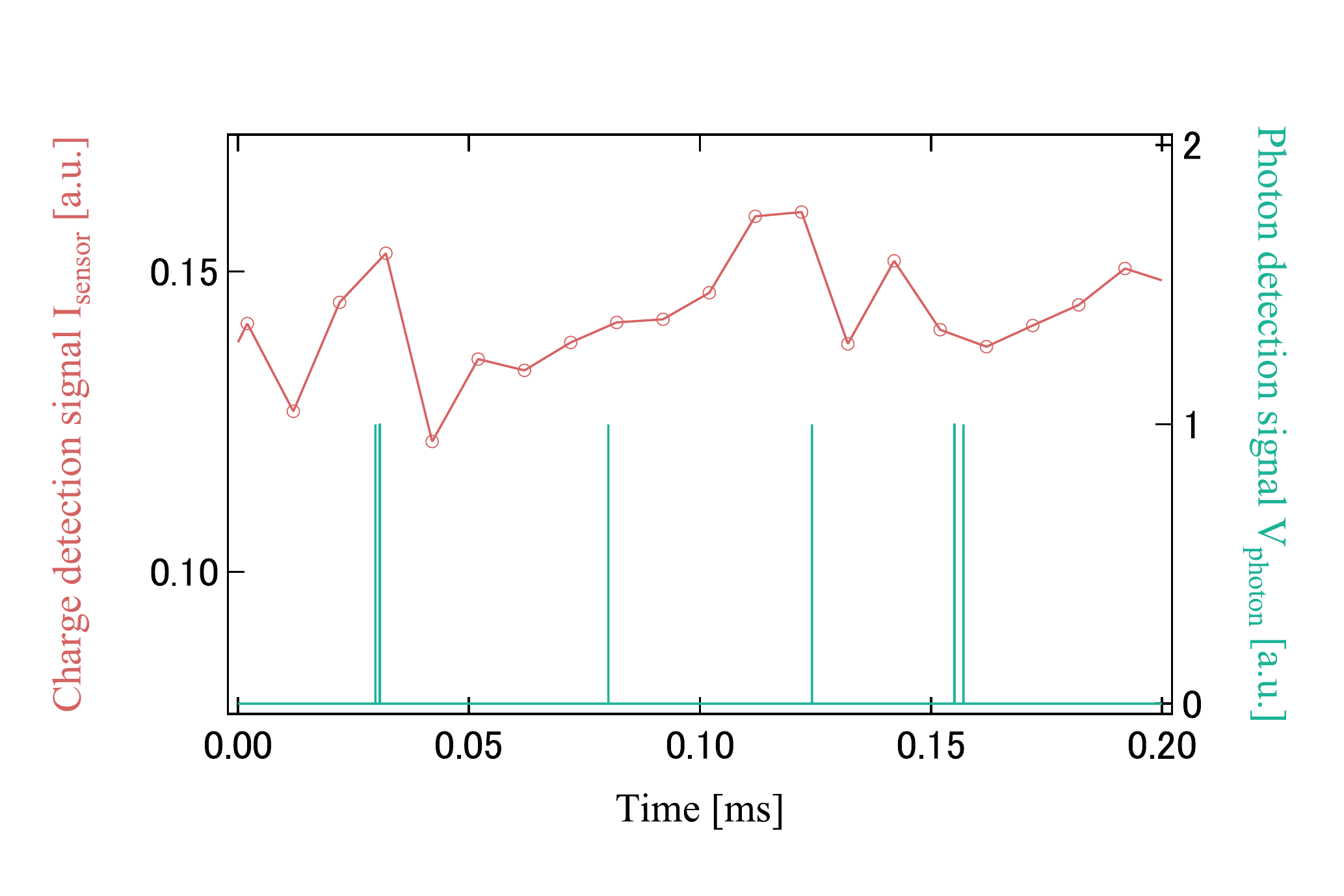}
\caption*{Fig.S2 An example time trace of the random samplings taken while the electron trapping dose not occur. We counts the number of the detected photons for every 10 $\rm{}\mu$sec interval of the charge sensor. In this time trace one or two photons are observed in 10 $\rm{}\mu$sec.}
%\label{figureS2}
\end{center}
\end{figure}
\newpage

%figure S2
%%
\begin{figure}[H]
\begin{center}
\includegraphics[width=10cm]{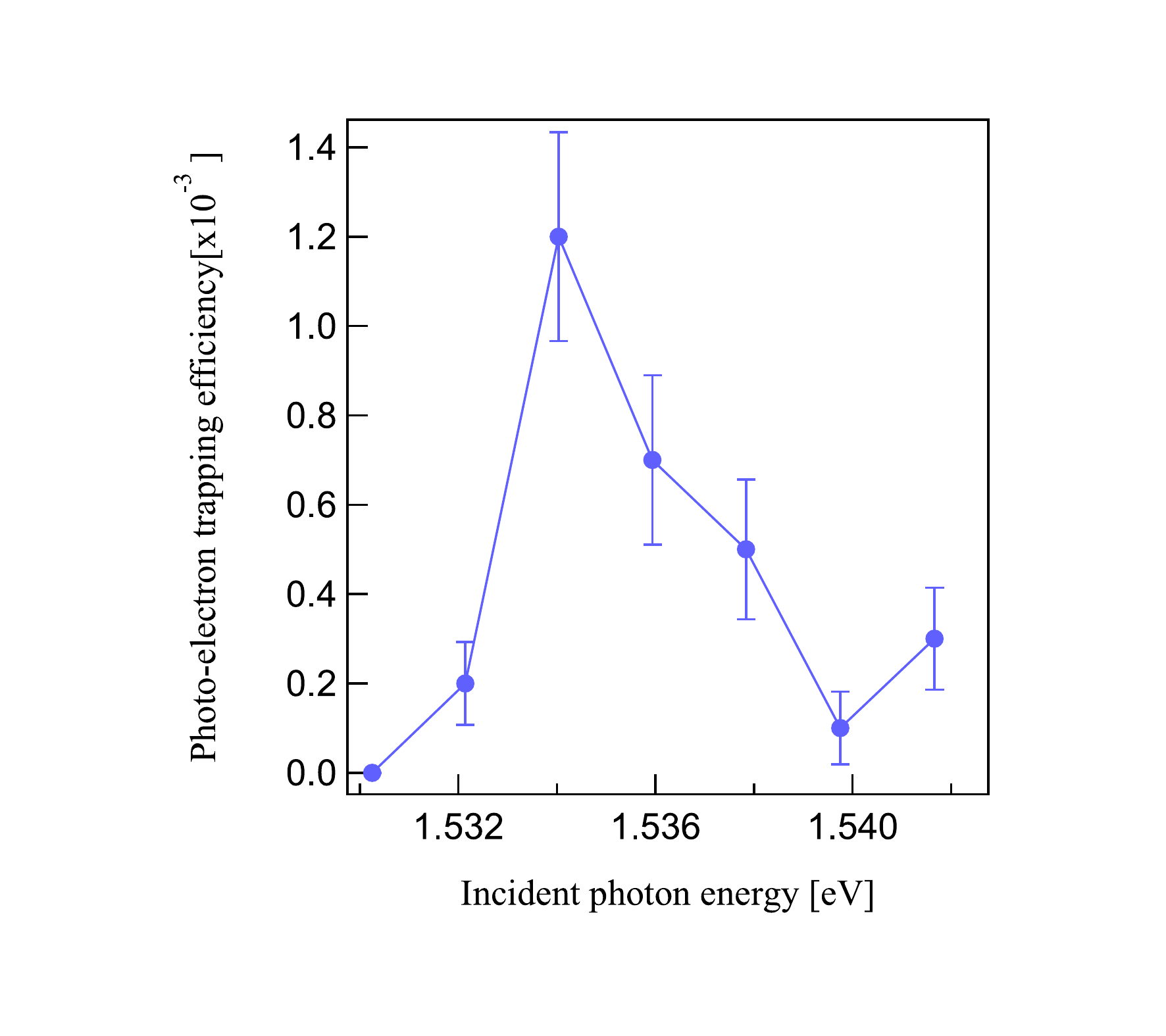}
\caption*{Fig.S3 Energy spectrum of photo-electron trapping efficiency on the QD taken under external in-plane magnetic field of 9 T. A horizontal axis is energy of an incident photon. A peak at 1.534 eV corresponds to resonant excitation of heavy hole exciton.}
%\label{figureS2}
\end{center}
\end{figure}
\newpage
% 参考文献リスト
\fancyhead{}
\rhead[]{reference}
\lhead[reference]{}
  %\addcontentsline{toc}{chapter}{\bibname}
%  \phantomsection
\begin{spacing}{1.1}
%\bibliographystyle{fujita-copy}
%\bibliographystyle{junsrt}
%\bibliography{C:/Users/kuroy_000/Desktop/paper_work/Tex/bibfile}

%\nocite{*}
\end{spacing}
\end{document}